\documentclass[12pt,english]{article}
\usepackage {epsfig}

\usepackage{babel}
\usepackage[usenames]{color}

\begin{document}
\baselineskip 18pt
\def\today{\ifcase\month\or
 January\or February\or March\or April\or May\or June\or
 July\or August\or September\or October\or November\or December\fi
 \space\number\day, \number\year}
\def\thebibliography#1{\section*{References\markboth
 {References}{References}}\list
 {[\arabic{enumi}]}{\settowidth\labelwidth{[#1]}
 \leftmargin\labelwidth
 \advance\leftmargin\labelsep
 \usecounter{enumi}}
 \def\newblock{\hskip .11em plus .33em minus .07em}
 \sloppy
 \sfcode`\.=1000\relax}
\let\endthebibliography=\endlist
\def\lsim{\ ^<\llap{$_\sim$}\ }
\def\gsim{\ ^>\llap{$_\sim$}\ }
\def\r2{\sqrt 2}
\def\beq{\begin{equation}}
\def\eeq{\end{equation}}
\def\beqn{\begin{eqnarray}}
\def\eeqn{\end{eqnarray}}
\def\rmuu{\gamma^{\mu}}
\def\rmud{\gamma_{\mu}}
\def\PL{{1-\gamma_5\over 2}}
\def\PR{{1+\gamma_5\over 2}}
\def\sinW2{\sin^2\theta_W}
\def\AEM{\alpha_{EM}}
\def\mul{M_{\tilde{u} L}^2}
\def\mur{M_{\tilde{u} R}^2}
\def\mdl{M_{\tilde{d} L}^2}
\def\mdr{M_{\tilde{d} R}^2}
\def\mz2{M_{z}^2}
\def\c2b{\cos 2\beta}
\def\au{A_u}
\def\ad{A_d}
\def\cob{\cot \beta}
\def\v#1{v_#1}
\def\tb{\tan\beta}
\def\epem{$e^+e^-$}
\def\KK{$K^0$-$\bar{K^0}$}
\def\wi{\omega_i}
\def\xj{\chi_j}
\def\Wmu{W_\mu}
\def\Wnu{W_\nu}
\def\m#1{{\tilde m}_#1}
\def\mH{m_H}
\def\mw#1{{\tilde m}_{\omega #1}}
\def\mx#1{{\tilde m}_{\chi^{0}_#1}}
\def\mc#1{{\tilde m}_{\chi^{+}_#1}}
\def\mwi{{\tilde m}_{\omega i}}
\def\mxi{{\tilde m}_{\chi^{0}_i}}
\def\mci{{\tilde m}_{\chi^{+}_i}}
\def\mz{M_z}
\def\sw{\sin\theta_W}
\def\cw{\cos\theta_W}
\def\cb{\cos\beta}
\def\sb{\sin\beta}
\def\rwi{r_{\omega i}}
\def\rxj{r_{\chi j}}
\def\rfp{r_f'}
\def\Kik{K_{ik}}
\def\Fq2{F_{2}(q^2)}
\def\f{\({\cal F}\)}
\def\d1{{\f(\tilde c;\tilde s;\tilde W)+ \f(\tilde c;\tilde \mu;\tilde W)}}
\def\tw{\tan\theta_W}
\def\sec2w{sec^2\theta_W}
\newcommand{\In}{\mbox{\large$\in$}}
\begin{titlepage}
\begin{flushright}
\date{\today}
\end{flushright}
\begin{center}
{\LARGE {\bf Tiger Tales: A Critical Examination of the Tiger's Enclosure at the San Francisco Zoo}}\\
\vskip 0.5 true cm \vspace{2cm}
\renewcommand{\thefootnote}
{\fnsymbol{footnote}}
 \large{\textsl{\textsf{Erica Walker}}$^{a}$, and \textsl{\textsf{Raza M.
 Syed}}$^{b,c}$}
\vskip 0.5 true cm
\end{center}

\noindent{$a$. Boston Architectural College, 320 Newbury Street, Boston, MA 02115 \\
$b$. Health Careers Academy, 110 The Fenway, Cahners Hall, Boston, MA 02115\\
{ $c$. Department of Physics, Northeastern University,
360 Huntington Ave., Boston, MA 02115-5000}} \\

\vskip 1.0 true cm
\centerline{\bf Abstract}
\medskip
\noindent

\noindent{\textsf{Given the recent tragedy involving a 350 pound Siberian Tiger and the death of teenager Carlos Souza Jr., one must ask a fundamental question: Can a tiger overcome an obstacle that is thirty-three feet away and twelve and a half feet tall? Are these dimensions sufficient enough to protect the zoo-visitors from a potential escape and/or attack?
To answer these questions we use simple two-dimensional projectile motion to find the minimum velocity a tiger needs in order to clear the obstacle. With our results we conclude that it is highly likely that the tiger was able to leap over the obstacle with ease!}}\\

\end{titlepage}

\noindent\textsf{We begin by first writing down the two-dimensional kinematical equations satisfied by the projectile (\textbf{tiger}). To that end, assume that the initial velocity of the tiger just before being airborne is $v_0$ at an angle $\theta$ with respect to the horizontal. See Figure 1. Then the horizontal and vertical displacements $x$ and $y$, respectively, covered by the tiger in time $t$ is given by
\begin{eqnarray}
\displaystyle x&=&(v_0\cos\theta) t\label{xmotion}\\
\displaystyle y&=&(v_0\sin\theta) t-\frac{1}{2}gt^2\label{ymotion}
\end{eqnarray}
\begin{figure}[h]
\centering
\epsfig{file=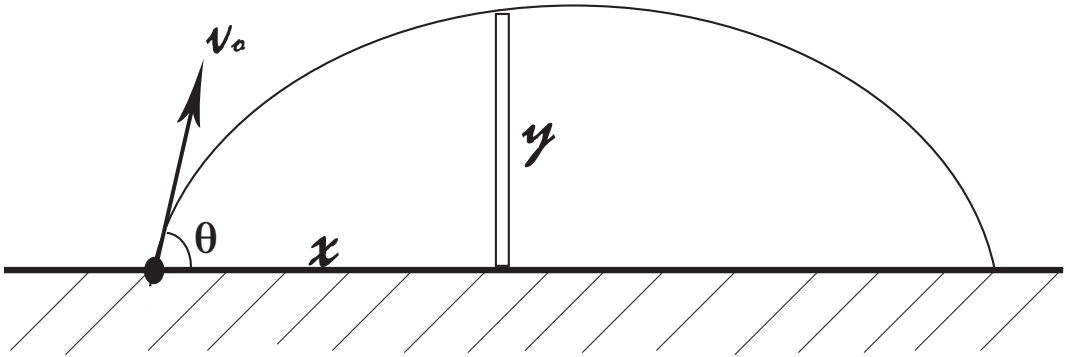, width=\textwidth}
\caption{ Variables Associated With the Tiger.}
\label{fig:example}
\end{figure}
Note that $g$ is acceleration due to gravity near the surface of the earth ($\approx 32~ft/s^2$) and displacements  are fixed quantities as made available by the San Francisco Zoo: $x=33~ft$ and $y=12.5~ft$.\\
\noindent Solving for $t$ in Eq.(\ref{xmotion}),
$$\displaystyle t=\frac{x}{v_0\cos\theta}$$
and substituting it in Eq.(\ref{ymotion}), we get
\begin{eqnarray}
\displaystyle y=x\tan\theta-\frac{1}{2}\frac{gx^2}{v_0^2}\textnormal{sec}^2\theta\label{par}
\end{eqnarray}
Solving for $v_0$ in this last equation, we obtain
\begin{eqnarray}
\displaystyle v_0&=&x\sqrt{\frac{g}{2(x\sin\theta\cos\theta-y\cos^2\theta)}}\nonumber\\
&=&x\sqrt{\frac{g}{(x\sin2\theta-y\cos2\theta-y)}}\label{v0}
\end{eqnarray}
We now look for the value of the angle of projection, $\theta$ which minimizes tiger's initial velocity, $v_0$. Differentiating $v_0$ with respect to $\theta$ in Eq.(\ref{v0}), yields
$$\displaystyle \frac{dv_0}{d\theta}=-x\left(x\cos2\theta+y\sin2\theta\right)\sqrt{\frac{g}{\left(x\sin2\theta-y\cos2\theta-y\right)^3}}$$
To find the local extremum, we set $\displaystyle \frac{dv_0}{d\theta}=0$. This implies
$$\displaystyle x\cos2\theta+y\sin2\theta=0~\Rightarrow~\theta=\frac{1}{2}\tan^{-1}\left(-\frac{x}{y}\right)$$
Using right angle trigonometry, one obtains $\displaystyle \sin2\theta=\frac{x}{\sqrt{x^2+y^2}}$ and
$\displaystyle \cos2\theta=\frac{-y}{\sqrt{x^2+y^2}}$. Substituting these expressions in Eq.(\ref{v0}) we obtain,
$$\displaystyle v_{0min}=x\sqrt{\frac{g}{\sqrt{x^2+y^2}-y}}$$
Rationalizing the denominator, we get the final form for the minimum value of $v_0$:
\begin{eqnarray}
\fbox{$\displaystyle v_{0min}=\sqrt{g\left(y+\sqrt{x^2+y^2}\right)}$}\label{final}
\end{eqnarray}
Substituting in the data, we get
$$\fbox{$\displaystyle v_{0min}\approx 39.1~ft/s\approx26.7~mi/hr$}$$
Tigers can reach speeds up to $35~mi/hr$. They can attain such speeds only over short distances (about 10 yards) during an attack [\texttt{http://www.wonderquest.com}].}\\

\noindent \textsf{Using simple physics, we have shown a highly possible solution as to how the tiger was able to escape its enclosure. From our calculations it was shown that a tiger only needs a little over $26~mi/hr$ to cross the $33~ft$ moat and clear the $12.5~ft$ high wall. From the current data that is available, a tiger can attain a maximum speed of $35~mi/hr$. Hence, the current dimensions of the enclosure are not enough to ensure that a tiger will not escape.}\\
\\
\noindent{\bf\large{Alternate Solution}}\\
\textsf{Here we employ non-calculus techniques to arrive at the same result depicted in Eq.(\ref{final}). To that end, we start with Eq.(\ref{par}). Using the trigonometric identity, $\textnormal{sec}^2\theta=1+\tan^2\theta$ and rearranging Eq.(\ref{par}), we obtain,
\begin{eqnarray}
\displaystyle \tan^2\theta-\left(\frac{2v_0^2}{gx}\right)\tan\theta+\frac{2v_0^2y}{gx^2}+1=0\label{angproj}
\end{eqnarray}
This equation is quadratic in $\tan\theta$. If this quadratic equation has two real roots then there are two possible angles of projection. If the roots are complex then there is no angle of projection which will cause the tiger to clear the moat and the wall. If the roots are real and equal  then there is a unique launch angle with the required property. In ballistics, this is sometimes referred to as {\textsc{the anti-aircraft gunnery problem}}.}

\textsf{For real and distinct roots, the discriminant of Eq.(\ref{angproj}) must be non-negative:
$$\displaystyle\left(\frac{-2v_0^2}{gx}\right)^2-4\left(\frac{2v_0^2y}{gx^2}+1\right)\geq 0
$$
Rearranging, we obtain a quartic inequality in $v_0$:
\begin{eqnarray}
\displaystyle v_0^4-\left(2gy\right)v_0^2-g^2x^2\geq 0\label{ineq}
\end{eqnarray}
The correct range of values for $v_0$ must be found by first obtaining the boundary points:
$$\displaystyle v_0=\sqrt{\frac{2gy\pm\sqrt{4g^2y^2+4g^2x^2}}{2}}=\sqrt{g\left(y\pm\sqrt{x^2+y^2}\right)}$$
Thus the range of values of $v_0$ for which the Inequality(\ref{ineq}) holds true, would be
$$\displaystyle v_0~\In~\left(-\infty,\sqrt{\displaystyle g\left(\displaystyle y-\sqrt{x^2+y^2}\right)}\right]\bigcup \left[\sqrt{g\left(y+\sqrt{x^2+y^2}\right)},+\infty\right)$$
However, since $y<\sqrt{x^2+y^2}$, the first region above will not be a valid one. In addition, if we assume the maximum velocity of tiger to be $35~mi/hr$ (see the reference on the previous page), then the correct solution to the above inequality would take the form:
$$\fbox{$\displaystyle v_0~\In \left[\sqrt{g\left(y+\sqrt{x^2+y^2}\right)},35~mi/hr\right]$}$$
\textsf{This is the same result as shown in Eq.(\ref{final})!}\\
An interesting result that is worth exploring are the launch angles  corresponding to a given initial velocity, $v_0$
that is necessary to achieve the known horizontal and vertical displacements, $x$ and $y$, respectively. Solving for $\theta$ in Eq.(\ref{angproj}), we get:
$$\displaystyle\theta=\tan^{-1}\left\{\displaystyle\frac{v_0^2\pm\sqrt{\displaystyle v_0^4-2gyv_0^2-g^2x^2}}{gx}\right\}$$
We now compute the values of $\theta$ using the given data ($x=33~ft,~y=12.5~ft$):
\begin{itemize}
\item $\displaystyle v_0=v_{0min}\approx 26.7~mi/hr,~\theta\approx55.4^0$
\item $\displaystyle v_0\approx 31~mi/hr,~\theta\approx38.5^0~ \& ~72.2^0$
\item $\displaystyle v_0=v_{0max}\approx 35~mi/hr,~\theta\approx33.7^0~ \& ~77.0^0$
\end{itemize}
Note that the values of $\theta$ are such that they are always symmetrical about the line $\displaystyle 45^0+\frac{1}{2}\alpha$, where
$\alpha$ is the angle of sight to the tip of the wall from the initial point of projection and it is given by
$\displaystyle\alpha=\tan^{-1}\left(\displaystyle\frac{12.5}{33}\right)\approx 20.7^0$.}
\end{document}